\documentclass[3p]{elsarticle}

\usepackage{amsmath}
\usepackage{amssymb}

\usepackage{url}

\journal{Physica A}

\begin{document}
\begin{frontmatter}

\title{Activist Model of Political Party Growth}
\author{Rebecca A. Jeffs}
\author{John Hayward\corref{cor1}}
\cortext[cor1]{Corresponding author}
\ead{john.hayward@southwales.ac.uk}

\author{Paul A. Roach \corref{cor2}}
\author{John Wyburn}

\address{School of Computing and Mathematics, University of South Wales, Pontypridd, CF37 1DL, Wales, UK}

\begin{abstract}
The membership of British political parties has a direct influence on their political effectiveness. This paper applies the mathematics of epidemiology to the analysis of the growth and decline of such memberships. The party members are divided into activists and inactive members, where all activists influence the quality of party recruitment, but only a subset of activists recruit and thus govern numerical growth. The activists recruit for only a limited period, which acts as a restriction on further party growth. This Limited Activist model is applied to post-war and recent memberships of the Labour, Scottish National and Conservative parties. The model reproduces  data trends, and relates realistically to historical narratives. It is concluded that the political parties analysed are not in danger of extinction but experience repeated periods of growth and decline in membership, albeit at lower numbers than in the past. 
\end{abstract}


\end{frontmatter}

\section{Introduction}
Political parties play a vital role in the governance of
countries. They provide the personnel out of which national and
local leaders emerge,  a legitimate political identity for those
in government, and an arena in which policy can be formed. Even when
not in power, parties can provide checks and balances by providing
an opposition, both in democratic forums and outside. As such
there is keen interest in their growth and decline.

Many theories of political party growth consider the relationship
between its membership size and its growth. One of the earliest such theories was 
Michels' ``Iron Law of Oligarchy'' which states that any form of
organisation will eventually develop an oligarchy as it grows to
the point where real democracy becomes difficult
\cite{Michels:Political}. The reasons given for this oligarchy are that
a large complex organisation requires a specialist bureaucracy in
order to make efficient day to day decisions, thus inevitably
removing the rank and file members from the centre of power.
Michels applied this theory, originally published in 1915, to the
growing socialist parties of Europe, showing that they evolved into
oligarchies as did earlier conservative parties, despite all their
ideals. This has clear implications for the growth of a party as
the membership suffers diminishing ability to
become involved in decision making, rendering membership less attractive.

\citet{Tan:size}  hypothesised that there
would be a direct effect of increasing party size on member
participation and growth due to free riding, that is becoming inactive, as proposed by
\citet{Olson:Logic}, as well as there being an indirect effect due to
increasing complexity, as put forward by
\citet{Michels:Political}. Using data on 23 political parties
\cite{Janda:Political}, Tan argued that although party size can
directly reduce participation, the indirect effect through
increasing complexity can positively influence the participation
of members, thus contradicting that aspect of Michel's law.
Although the debate as to the effect of party size on participation
is still not resolved \citep[e.g.][]
{Seyd:British,Weldon:polity,Whiteley:Dynamics} it is clear
that there is a potential limit to the size of a political party
related to its ability to keep members active. This suggests the
importance of activists in the growth of the party.

The importance of activists in the interests of a political party has been
emphasised by \citet{Seyd:British}, noting their role in fund
raising, political legitimacy, and as a source of voters,
campaigners and potential candidates for parties. However they
argue that party decline is due to choice rather than the
structural reasons of \citet{Tan:size} and
\citet{Michels:Political}. They further claim that a party
chooses to restrict the supply of activists as they can also be a
hindrance to the party leadership's freedom of action. The
reduction in  the benefits for party membership, and the restriction of
activists, is thus deemed to be one of the causes of the decline.
 In this case party growth comes from the often brief periods during which a party requires activists for electoral purposes.

Further support for belief in the importance of the role of activists in party growth is
provided by \citet{Weldon:polity} and
\citet{Norris:Phoenix} who both showed that small parties
have a higher degree of activism compared with larger parties due
to the lack of funds to support a professional organisational
structure. It is thus suggested that larger parties find it harder
to grow, as the reduced incentives that can be offered are insufficient to keep activists in the
party or to maintain their level of activism.

However a decline in party size does not necessarily increase
activism as the party can remain organisationally complex if it lacks the funds to support the change back to a state in which
activism could again flourish \cite{Tan:decline}. By contrast some
parties can develop  complex structures to make
organisation and activity more efficient and effective
\cite{Tan:size}. These phenomena suggest that activism is a natural feature of the smaller and growing party, but not of the larger and diminishing. However, larger parties are capable of sustaining activism, at least during those periods where activism is not contrary to the oligarchy's aims. In either case, activism is the cause of growth.

It is inevitable in the light of the above that  much of the
attention on change in party size concentrates  on  decline
\cite{Whiteley:Dynamics,Tan:decline,Dalton:Partisans,Whiteley:High},
with less attention being given to the mechanisms behind their
growth. Discussions on growth generally focus on
the reasons \emph{why} parties wish to grow and the incentives provided
for such increases in membership
\cite{Seyd:British,Weldon:polity}. There has been
less discussion on \emph{how} parties grow but it is clear
that activists, those most involved in the life of the party, are
key to its growth \cite{Norris:Phoenix,Whiteley:High}.

The question needs to be asked as to how activists recruit new
members to the party. \citet{Jeffreys:History} points to very
specific recruitment campaigns that dramatically increased the
membership of the main UK political parties in the immediate
post-war period. These campaigns were largely carried out using door-to-door recruitment by the most active members of the parties. Also a
deliberate recruitment strategy by the Labour party from 1994
resulted in it temporarily becoming  the fastest growing party in
Europe \cite[p. 24]{Whiteley:High}. These periods of growth could be explained in terms of a word-of-mouth  phenomenon driven by party activists.

This paper proposes a word-of-mouth model of political party growth using ideas similar to mathematical epidemiology. The model divides the party into activists and inactive members. Activists are further divided into those who recruit (the ``infectious'') and those who do not recruit. However non-recruiting activists do contribute to the party by discouraging new members from free riding. Thresholds of growth are computed and the results of the model are compared with historical party data from the UK. 

\section{Previous Models of Social Diffusion}
Word-of-mouth models for social diffusion and organisational growth are not new. \citet{Burbeck:Riot} used an SIR model to investigate the spread of rioting, applying the model to three riots from the 1960s. Similar models with more variables have been applied to riots, public outrage  and terrorist groups \cite{hayward2014model,nizamani2013public,camacho2013development}. 

Other models dealing with the spread of behaviour include language acquisition \cite{Abrams:language, wyburn2008future}, alcohol consumption \cite{Manthey:drink,Sanchez:Drink}, cigarette smoking \cite{Rowe:Smoke} and psychological/social diseases such as bulimia and obesity \cite{gonzalez2003too,santonja2010mathematical}.  Most of these models employ multiple sub-populations with one or more acting as infectious agents and at least one non-infectious. In contrast, the influential Abram-Strogatz model \cite{Abrams:language} assumes there is no non-infectious category, a specific feature of language acquisition, making it less applicable to the other  behaviours modelled.

Additionally epidemiological ideas have been applied to the spread of rumours \cite{Kawachi:Rumour,zhao2011sihr}, ideologies  \cite{santonja2008mathematical,vitanov2010verhulst} and online networks \cite{cannarella2014epidemiological,woo2013modeling,chhabra2014alternative}. \citet{bettencourt2006power} modelled the spread in the use of Feynman diagrams throughout the scientific community. Their model allowed for  some new recruits to the Feynman methodology to be non-infectious, an exposed class, thus allowing for a weaker growth in recruiters compared with all users of the methodology.

None of the above models involve  organisational membership as seen in political parties. However models close to these ideas have been used for the spread of religious affiliation.  \citet{Hayward:Church,Hayward:General} applied SIR type models to church and denominational membership where the infectious church members responsible for recruitment, called ``enthusiasts'', were a subset of new recruits. The models were applied to a range of religious denominations which could be categorised as experiencing rapid growth, as stable or as heading for extinction. Like  \citet{bettencourt2006power}, \citet{Hayward:General} also allowed for some new recruits to be non-infectious. By contrast \citet{ochoche2013evaluating,madubueze2014mathematical} considered the longest serving church members as the recruiters.  \citet{mccartney2015three} modelled a variety of recruitment mechanisms, predicting that a range of denominations in Northern Ireland faced future extinction.

Additionally there are  statistical physics models of religious adherence
\cite{ausloos2007statistical, ausloos2009statistical,ausloos2012econophysics} that employ the epidemiological analogy.
In particular \citet{Ausloos:econ, abrams2011dynamics} draw the parallel between the spread of language and religion. The model of \citet{abrams2011dynamics} has the same restriction of no inactive members as the Abrams-Strogatz language acquisition model, on which it is based.

One specific epidemiological model related to political party growth  was presented by \citet{romero2009epidemiological} who modelled the rise of third political parties in a state dominated by two parties. In the model the third party seeks to recruit by changing voter opinion as  small start-up parties do not have a voter base from which to recruit. Results illustrate the critical role activists play in sustaining grassroots movements and indentify conditions under which such minority parties can thrive. Although primarily about opinion change, and not directly relevant to the growth of mainline political parties, their model will provide a useful comparison in model construction. Additionally there are opinion dynamics models that link political party size to party characteristics \cite{galam2000dictatorship}, and that model the influence of party activists on opinion change \cite{qian2015activeness}.

Although not exhaustive, the above review is sufficient to justify word-of mouth-modelling being applied to political party growth and decline.

\section{Initial Considerations}
To investigate the hypothesis that political party growth is due to the
recruitment activities of party
activists,  the SIR epidemic model is taken as an initial exploratory model. This leads to the following assumptions:

\begin{description}
    \item[Assumption 1]
The population is partitioned into susceptibles $S$, those who do not
belong to the political party, and party members $P$. Party members are
further partitioned into activists $A$ and inactive members $M$.
This latter group are only inactive in the sense of recruitment.
They may still attend party functions and engage in other activities.
    \item[Assumption 2]  The activists take on the role of the infectives in a disease and the
model assumes that they only remain active for an average time
$\tau$.

    \item[Assumption 3]  All new recruits become
activists with the same average ability to recruit others to the party.

\item[Assumption 4] The contacts that activists make are limited by
how many people they can meet in a given time period, rather than
by the density of the population in which they work. This is more in
keeping with deliberate door-to-door recruitment campaigns and other
word-of-mouth phenomena in a large population. Thus the standard  incidence model of
contacts is assumed, similar to that used in the spread of sexually
transmitted diseases \cite{hethcote:effects}. 
\end{description}
The model is intended to be applied to short time periods only, thus births, deaths, leaving and migration are ignored. Thus the equations are:
\begin {eqnarray}
\frac {dS} {dt}  &=&-\frac{C_{p}}{\tau N} SA \label {standard.1} \\
\frac {dA} {dt}  &=&\frac{C_{p}}{\tau N} SA  - \frac{A}{\tau}
\label {standard.2} \\
\frac {dM} {dt}  &=& \frac{A}{\tau}  \label{standard.3}
\end {eqnarray}
where $N=S+A+M$ is the total adult population, a constant; homogeneous mixing has been assumed. $C_{p}$ is  the recruitment potential, that
is the number of susceptibles recruited, or converted, to the
party by one activist  throughout their entire active period.

The condition for epidemic growth in the
party can be derived from setting $\dot{A}>0$ in (\ref{standard.2}), giving $S/N>1/C_{p}$. This can be re-arranged to provide a  threshold value for $C_p$ over which an epidemic will occur:
$ C_{p} > 1/(S/N) \triangleq C_{\mathrm{epi}}$. 

In this standard incidence SIR model the epidemic threshold $C_{\mathrm{epi}}$ depends on the fraction of susceptibles alone. The larger the fraction susceptible, the lower the threshold and the less recruitment potential to give epidemic party growth, as also seen in models of disease \cite{anderson1992infectious}.   Thus rapid growth becomes harder to achieve as the pool of susceptibles reduces, and party growth will cease because
activists are no longer  able to reproduce themselves sufficiently quickly to make up for their losses. Growth ends before all people
join the party.

Importantly the threshold $C_{\mathrm{epi}}$ does not
depend on the number of activists. One activist is sufficient to start
epidemic growth if the recruitment potential is sufficiently high (at
least $C_{p}>1$ ) and the proportion of susceptibles in the population
is sufficient. This assumes activists can reproduce activists from
their recruits with the same ability to recruit. In this case $C_{p}$ is the equivalent of  the
basic reproduction number $R_{0}$ for the epidemic \cite{hethcote:effects, anderson1992infectious}.

To test the SIR model, data from the UK Labour party are used from 1947--1953 (Table \ref{lab44.tab}), a period of rapid growth  resulting from a deliberate door-to-door recruitment campaign by the party with the intention of making every Labour voter a party member, \citep[][pp.~93--99]{Jeffreys:History}. The party membership, $P=A+M$,  in the model (\ref{standard.1}--\ref{standard.3}) is compared with the data using a Runge-Kutta-Fehlberg method and optimised using least squares. Because of the uncertainty in the fraction of the adult population reached by activists during the period, three optimised runs are produced:  (i)~$N=33.24$ million, the electorate in the 1945 election, those with a known interest in politics \cite{uk:political}; (ii)~$N=11.995$ million, the number of Labour voters in the 1945 general election, the people worth an activist spending time with \cite{uk:political}; and (iii)~$N=5$ million, a more conservative estimate based on average pre-war union membership, the hard core Labour supporters \cite{keen2014membership}. It is unclear the extent to which the recruitment campaign was targeted towards likely recruits through, for example, the trade union movement.

The model parameters estimated for the three population sizes, (i)--(iii),  are respectively $C_p=$ 1.025, 1.073, 1.194 ; $\tau = 1.1939, 0.0100, 0.0283$ and $A_0/P_0= 0.005\%,$ $ 0.011\%, 0.040\%$. In each case the  threshold $C_{\mathrm{epi}}$ exceeds the recruitment potential from about 1949, indicating the slowing down of party growth. Irrespective of the target population size, the durations spent as activist, $\tau$, are very small, merely a matter of weeks at best. While this may be an acceptable duration for a spreader in word-of-mouth diffusion, it is far too short a period for a political activist, some of whom remain active for many years \cite{Jeffreys:History}. Such a short duration leads to an unacceptably small number of activists, less than 1\% of the party at the peak. Estimates of party activists for major political parties range from 30-40\%  \citep[][ch.~12]{Janda:Political} \citep[][p.~127]{Norris:Phoenix}. Thus the SIR model is not suitable as a model of political party growth without modification.

It is proposed that the fault in the modelling  lies with the identification of recruiters with activists. Recruitment is only a small part of an activist's role in the party; most of their time is spent on campaigning, fund raising and political discussion \cite{Weldon:polity, Norris:Phoenix,Whiteley:High}. \citet{pedersen2004sleeping} found that only 10\% of activists spent more than 5 hours a month on recruitment and electoral duties, indicating that recruitment is very much a minority activity of party activists. Thus it is proposed in the next section that the model be extended to allow for recruiting and non-recruiting activists, giving the latter a distinct dynamical role in the quality, rather than the quantity, of recruitment.

\section{Limited Activist Model Construction}
The model of Section 3 is modified to distinguish between activists who recruit and those who do not. All activists have a role in determining the quality of the party as measured by the proportion who are active. Further the issue as to what fraction of the population is genuinely open to joining the party is resolved by introducing a separate population of people hardened against the party beliefs. The revised model will be referred to as the Limited Activist  model.
\begin{description}
    \item[Assumption 1]
The population is partitioned into party members $P$ and non-party members.  Party members are
further partitioned into recruiting activists $I$, non-recruiting activists $A$ and inactive members $M$.
This latter group are free-riders who do not participate in any party activities. The non-recruiting activists contribute to the political activity of the party but are ineffective in recruitment. The non-party members are partitioned into susceptibles $S$, those open to the party because of their political persuasion, and hardened $H$, those sufficiently opposed to the party that they would never become members.
    \item[Assumption 2]  The recruiting activists take on the role of the infectives in the spread of a disease and become non-recruiting activists after an average time $\tau_i$. Non-recruiting activists become inactive after an average time $\tau_a$. It is expected $\tau_a \gg \tau_i$ as recruitment is only carried out by a small minority of activists at any one time.

    \item[Assumption 3]  All new recruits may become either type of activist, or inactive members.

\item[Assumption 4] The contacts that activists make are limited by
how many people they can meet in a given time period, rather than
by the density of the population in which they work. 
\item[Assumption 5] All activists influence the number of recruits who themselves become active according to the proportion of activists in the party. That is a more politically active party will have more politically active recruits.
\item[Assumption 6] Inactive members leave the party at a fixed rate $\alpha$ and are open to rejoining, reflecting the ease with which members may lapse through non-renewal of membership.
\end{description}

Following from the above assumptions, the equations of the Limited Activist Model  are:
\begin {eqnarray}
\frac {dS} {dt}  &=&-\frac{C_p}{\tau_i N}SI + \alpha M  \label {model.S} \\ 
\frac{dI}{dt} &=& g\frac{C_p}{\tau_i N}SI        - \frac{I}{\tau_i}  \label{model.I} \\
\frac {dA} {dt}  &=&  f(1-g)\frac{C_p}{\tau_i N}SI + \frac{I}{\tau_i} -\frac{A}{\tau_{a}}\label{model.A}\\
\frac {dM} {dt}  &=&  (1-f)(1-g)\frac{C_p}{\tau_i N}SI  +\frac{A}{\tau_{a}}   - \alpha M \label{model.M} 
\end {eqnarray}
where
\begin{equation}
f=f(A_T) \triangleq \frac{I+A}{P} \label{model.f}
\end{equation} 
represents the fraction of those new recruits who do not become recruiters, but who nevertheless contribute to the party activism rather than free-riding. The coefficient $g$ is the fraction of recruits who become recruiters, $C_p$ is the recruitment potential, and total active membership $A_T=I+A$.

The total population $N=H+S+I +A+M$; total open population with sympathy for the party  $O=N-H$; and total party membership  $P=I +A+M$.  As the model is intended only for short-term growth and decline, births and deaths are excluded and the hardened $H$ and total population $N$ are assumed constant. Thus $(S,I,A,M)$ are determined by (\ref{model.S}--\ref{model.M}) with the stated constraints.

Multiple influential populations have been used in structurally similar models of the diffusion of bulimia \cite{gonzalez2003too}, binge drinking \cite{Manthey:drink}, rumours \cite{zhao2011sihr},  public violence \cite{nizamani2013public}, terrorist groups \cite{camacho2013development} and political opinion change \cite{romero2009epidemiological}.  This mechanism is required to allow for party members in $I$ and $A$ to have different types of activity; in particular many party activists do not recruit \cite{Weldon:polity,Norris:Phoenix,pedersen2004sleeping}. The Limited Activist model proposes that the influence of non-recruiting activists is on the quality of recruits, determined by $f(A_T)$, rather than on the quantity as in epidemiological mass action. However this does increase the number of activists at the expense of the inactive. \citet{galam2000dictatorship} have also modelled the influence of party size on the qualitative characteristics of a party, especially the role of activists in that process \cite{qian2015activeness}.

Multiple routes from susceptibles into different populations are used in  the  diffusion of scientific ideas \cite{bettencourt2006power}, rumours \cite{zhao2011sihr}, religion \cite{Hayward:General} and terrorist groups \cite{camacho2013development}.  Likewise \citet{romero2009epidemiological} have constant proportions entering different voter populations. What is different in the proposed Limited Activist model is that, rather than the proportions of recruits entering the three party member categories being fixed, the proportion becoming activists,  $f(A_T)$,  can vary, allowing a highly active party to become  more active, even if numerical recruitment is moderate. This is required to account for the differing effects of recruitment by parties with different levels of activism. For example \citet[][p.~126]{Whiteley:High} note the poor quality of  recruits in the Labour party in the 1990s, when the party was relatively inactive compared with campaigns in previous generations where activity, not just numbers, was improved. 

Multiple susceptible populations with different degrees of susceptibility were proposed by \citet{granovetter1978threshold} and have been used in the diffusion of binge drinking \cite{Manthey:drink}, religion \cite{Hayward:General}, terrorist groups \cite{camacho2013development}, riots \cite{hayward2014model}  and knowledge diffusion  \cite{ausloos2015slow}. The Limited Activist model proposes two such populations, one aligned to the party, $S$, and one opposed, $H$.  It is assumed that the latter  would not change political alignment and join the party on the timescales being considered. 

\section{Analysis}
As with the SIR model the epidemic threshold  follows from $\dot{I}>0$. Let  $R_p = g C_p$, the reproduction potential, or basic reproductive ratio, which measures how many recruiting activists one recruiting activist reproduces during their recruiting period, assuming the whole population $N$ is susceptible. Then from (\ref{model.I})
 \begin{equation}
   R_{p} >  \frac{1}{\bar{S}} \triangleq R_{\mathrm{epi}}
    \label{LAMthresh.2}
\end{equation}
where $R_{\mathrm{epi}}$ is the threshold value for the reproduction potential, above which there is an epidemic of recruiting activists. Barred variables denote the fraction of the total population $N$: $\bar{S} = S/N$.

Setting (\ref{model.I}) to zero gives either $I=0$ or $\bar{S} = 1/R_p$. The former is the equivalent of the disease-free equilibrium (DFE) where the party vanishes $(1-\bar{H}, 0,0,0)$.
 
The other equilibria will be determined from $\bar{S} = 1/R_p$. Applying this to (\ref{model.S}) set to zero,  with $\bar{A} = 1 - \bar{H}-\bar{S}-\bar{I} -\bar{M}$ gives:
\begin{equation}
\bar{A}= \hat{p}-\hat{q}\bar{I}   \label{model.Seq}
\end{equation}
where
\begin{eqnarray}
\hat{p} &=& 1 - \bar{H}-\frac{1}{R_p} \label{model.p} \\ 
\hat{q} &=&   \frac{g\alpha \tau_i + 1}{g\alpha \tau_i} \label{model.q} .
\end{eqnarray}
Further,  $f = 1-\bar{I}/(g\alpha \tau_i \hat{p})$, using 
$ \hat{p} = \bar{I} +\bar{A}+\bar{M}$. Thus 
a single equation for  $\bar{I}$ can be obtained by setting (\ref{model.A}) to zero, and the substitution of $f$  and $\bar{A}$ (\ref{model.Seq}), giving:
\begin{equation}
a_1 \bar{I}^2 - b_1 \bar{I} + c_1 = 0 \label{model.a1} 
\end{equation} 
with
\begin{eqnarray}
a_1 &=& \frac{(1-g)}{g^2\alpha \tau_i^2 } \label{model.a2} \\
b_1 &=& \hat{p}\left[\frac{1}{g\tau_i} + \frac{\hat{q}}{\tau_a}\right] \label{model.a3} \\
c_1 &=&  \frac{\hat{p}^2}{\tau_a} \label{model.a4} 
\end{eqnarray}
Thus
\begin{equation}
\bar{I}_{\pm} = \frac{b_1 \pm \sqrt{ b_1^2 - 4a_1c_1}}{2a_1} \label{model.a5}
\end{equation} 

Note that $a_1$ and $c_1$ are always positive and the sign of $b_1$ depends on the sign of $\hat{p}$, as $\hat{q}$ is positive.
Thus there are two endemic equilibria (EE) with differing recruiting activists, $\bar{I}_{\pm}$ but the same party size $P=N-H-S$, given by $(1/R_p,I_{\pm},\hat{p}-\hat{q}\bar{I}_{\pm},\bar{I}_{\pm}/(g\alpha \tau_i))$. Both roots of (\ref{model.a1}) are real as its discriminant reduces to:
$$\left[\tau_a-\frac{1}{\alpha} \right]^2+ 2\tau_a g\tau_i+g^2\tau_i^2+2\frac{g\tau_i}{\alpha}  + \frac{4g\tau_a}{\alpha  } 
$$
which is always positive (\ref{model.a2}--\ref{model.a4}). Neither root is physical when $\hat{p}<0$, as from (\ref{model.a5}) $\bar{I}_{\pm}<0$. However the one root is never physical as $A_{+}=\hat{p}-\hat{q}\bar{I}_{+}$ is always negative (\ref{proof.appendix}). Thus there are only two relevant equilibrium points, the DFE and a single EE given by $I_{-}$ in (\ref{model.a5}), the latter only physical when $\hat{p}>0$.

Stability of the two remaining equilibria is deduced from the Jacobian of (\ref{model.S}--\ref{model.M}):
\begin{equation}
  J =
  \left[
  \begin{array}{cccc}
    -\frac{C_{p}\bar{I}}{\tau_i} & -\frac{C_{p}\bar{S}}{\tau_i}                     &    0    & \alpha\\
    \frac{gC_{p}\bar{I}}{\tau_i} &
\frac{gC_{p}\bar{S}}{\tau_i}-\frac{1}{\tau_i}  &
0             & 0\\
     f(1-g)\frac{C_{p}\bar{I}}{\tau_i}   &
f(1-g)\frac{C_{p}\bar{S}}{\tau_i} +\frac{1}{\tau_i}  +f_{,\bar{I}} Q   &-\frac{1}{\tau_a} +f_{,\bar{A}} Q & f_{,\bar{M}} Q\\
    (1-f)(1-g)\frac{C_{p}\bar{I}}{\tau_i}               & (1-f)(1-g)\frac{C_{p}\bar{S}}{\tau_i}  -f_{,\bar{I}} Q & \frac{1}{\tau_a} -f_{,\bar{A}} Q &- \alpha- f_{,\bar{M}} Q
  \end{array}
  \right]
  \label{model.j1}
\end{equation}
where  $f_{,\bar{I}} \triangleq \partial f / \partial \bar{I}$ etc, and
$Q = (1-g)C_p\bar{S}\bar{I} $.

For the DFE  the eigenvalues $\lambda$ of  (\ref{model.j1}) are:
\begin{equation}
\lambda \left(\frac{gC_{p}\bar{S}}{\tau_i}-\frac{1}{\tau_i}-\lambda\right) \left(\frac{1}{\tau_a} +\lambda\right) \left(\alpha +\lambda\right) = 0 \label{model.j5}.
\end{equation}

\noindent One eigenvalue is zero, as expected for a conserved system, and two are always negative. Thus stability of the DFE is determined by the remaining eigenvalue. As $\bar{S}=1-\bar{H}$ in equilibrium then the DFE is stable when:
$ R_{p}(1-\bar{H})/\tau_i < 1/\tau_i$
or
 \begin{equation}
R_{p}<\frac{1}{1-\bar{H}} \triangleq R_{\mathrm{ext}}\label{model.j6}
\end{equation} 
where $R_{\mathrm{ext}}$ is termed the extinction threshold, and unstable when $R_p > R_{\mathrm{ext}}$. If the reproduction potential is less than this threshold the party becomes extinct. If above the threshold the party will survive and reach the EE, with epidemic growth first if  $R_p$ is above the epidemic threshold. Note the extinction threshold is always below the epidemic threshold: $R_{\mathrm{ext}}\le R_{\mathrm{epi}}$.

When the DFE is stable,  $R_p < R_{\mathrm{ext}}$ and then, from (\ref{model.p}),  $\hat{p}<0$ and, as noted earlier, the EE is not physical. By contrast when the reproduction potential is above the extinction threshold then  the DFE is unstable,  $\hat{p}>0$, and thus the EE is physical and the party settles on a non-zero equilibrium. Thus the model displays a similar transcritical bifurcation as occurs in an SIRS model.

\section {Results}
The model is applied to three periods of rapid political party growth in the UK: the  Labour Party from post-war and in the 1990s, and the Scottish National Party (SNP) post 2000. In addition the model is applied to the recent decline in the membership of the UK Conservative Party.

\subsection{UK Labour Party 1944--1953}
The data, Table \ref{lab44.tab}, may be considered in three phases according to the change in numbers and historical evidence \cite{Jeffreys:History}. 

Firstly, the rise in membership 1944--1945 is due to the renewal of former members who had disengaged during the Second World War.

The second phase is associated with the short recruitment campaign of 1945--46. This was mainly conducted door to door, and followed on from the enthusiasm due to the Labour general election victory of 1945. This phase is assumed to have involved a high initial number of activists, including recruiting activists.   To estimate the total number of activists it is assumed the mid-war low of 219,000 \cite{keen2014membership} is a lower limit to the number of activists, those keen enough to retain membership during the war. Assume further that some activists re-joined in the first phase. Thus it is estimated that the second phase started with about 250,000 activists. 

The  Limited Activist model was applied to the rise in data 1945--46 from 487,000 to 645,000, and the  post campaign decline to 608,000 in 1947. The most convincing data fit  has the reproduction potential under the epidemic threshold with the growth due to a high initial number of recruiting activists of about $I_0=4000$.  This leaves about 200,000 total activists in 1947, which corresponds to roughly a third of the party active, as also reported by \citet{Janda:Political} and \citet{Norris:Phoenix}. This figure is used as the initial value in the third phase of growth, which is examined in greater detail.  

The third phase of growth in Labour party membership ran from 1947--1953 and was a more measured campaign conducted through the Trade Union movement as well as door to door. The UK electorate is used as the total population $N$, which is close to the adult population. The total open population $O=S+I+A+M$ is taken from those who voted Labour in 1945, i.e. those already members or open to joining due to their political persuasion. The time as a non-recruiting activist, $\tau_a$ is assumed to be 10 years. This is supported by \citet{Jeffreys:History}, who notes that a typical activist will be involved in at least one electoral campaign, implying up to 5 years as an activist. However the average figure will be higher than this as a number of activists remain so for life. The leaving rate $\alpha$ is taken as 5\% consistent with the average post-campaign decline to 1960  of 3\% per annum. No attempt to fit values 1954--1960 was made as the high volatility in change was assumed to be due to political factors for which a differential equation can only model the average. However, the 5\% leaving rate ensured the model was close to the 1961 figure of 751,000. This figure was the last year of decline in this phase \cite{keen2014membership}.   

The remaining parameters were obtained by least squares with a value for the fraction of recruits who become recruiters $g$ of 0.5, consistent with \citet{Jeffreys:History} and ensuring a high ratio of activist to inactive in the party. The 1951 data point was ignored as Labour lost an election that year and it is assumed the sudden drop was  a temporary reaction, followed by rejoining in 1952. The parameter values are given in Table \ref{lab2.tab}.

\begin{table}[!ht] \footnotesize
\begin{center}
\begin{tabular} {lll}
  \hline
Estimated  Parameters & Value & Source \\
  \hline
Total  population $N$ & 33.24 mill  & UK electorate 1945 election \cite{uk:political}  \\
Open population $O=1-H$  &  11.995 mill & Labour voters 1945 election \cite{uk:political}\\
Initial party membership $P_0$ & 608,000& Party data \cite{keen2014membership} \\
Initial activists $ A_{T0}=I_0+A_0$ &200,000 & Estimate from mid-war low and \\
&& roughly a third of party active \cite{Janda:Political,Norris:Phoenix}  \\
Duration non recruiting activist $\tau_{a}$ &10 years & Estimate anecdotal evidence \cite{Jeffreys:History}\\
Leaving rate $\alpha$&0.05& Estimate from post 1952 decline \cite{keen2014membership} \\
Fraction infected recruited $g$ & 0.5& High ratio of active to inactive recruits \\ 
Recruitment potential $C_{p}$&5.9501 & \emph{Optimised}\\
Duration recruiting activist $\tau_{i}$& 0.0144 & \emph{Optimised}  \\
Initial recruiting Activists $ I_0$ &  139  & \emph{Optimised} \\
  \hline
\end{tabular}
\end{center}
\vspace{-10pt}
\caption{Parameter Values for Limited Activist Model of UK Labour Party Membership 1947--1953}
\label{lab2.tab}
\end{table}

\begin{figure}[!ht]
\begin{center}
\includegraphics[height=10cm]{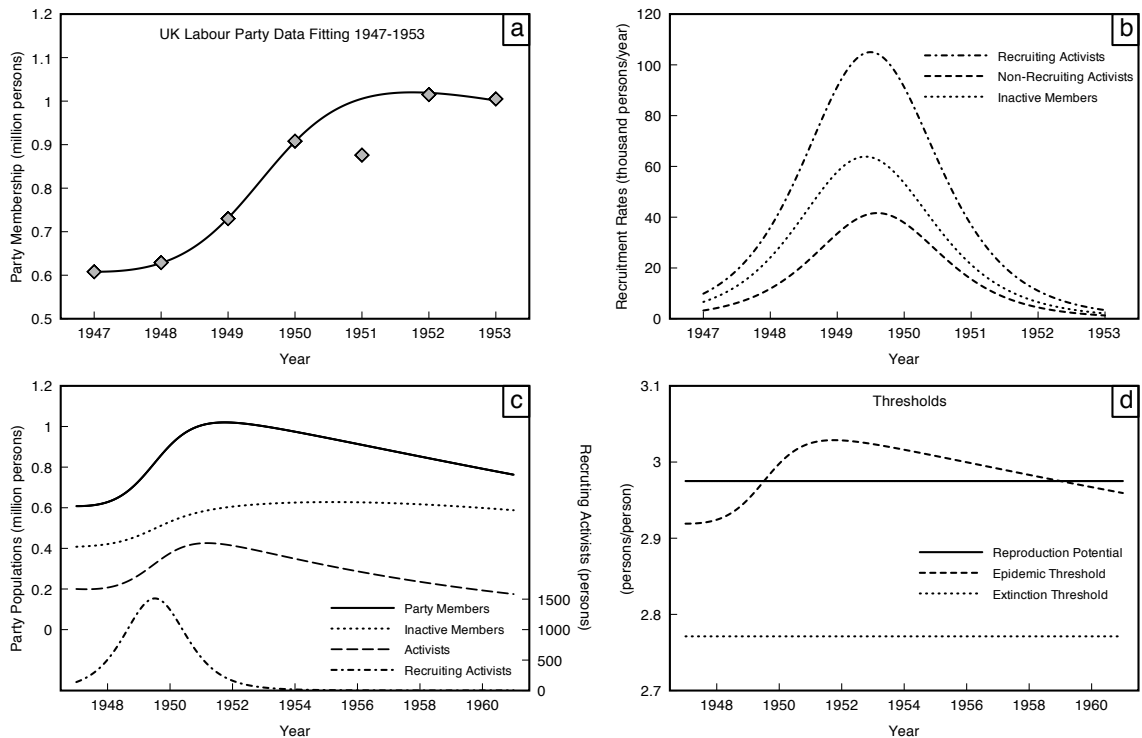}
\end{center}
\vspace{-18 pt}
\caption{Limited Activist Model applied to UK Labour Party. (a) Data fitting 1947--1953; (b) Recruitment rates to  party membership sub-populations $I$, $A$, $M$; (c) Party membership $P$ and sub-populations $I$, $I+A$, $M$; (d) Thresholds  $R_{\mathrm{epi}}$, $R_{\mathrm{ext}}$, compared with reproduction potential $R_p$.}
\label{politicalfig1.fig}
\end{figure}

A good fit is made to party data 1947--53, with the exception of 1951, a year with poor election results for Labour, Fig. \ref{politicalfig1.fig}(a). The largest share of recruitment is due to the recruiting activists, reflecting the known highly active nature of the party, Fig. \ref{politicalfig1.fig}(b). Although the recruiting activists peak at about 1949, Fig. \ref{politicalfig1.fig}(c), the total number of activists does not peak until two years later, rising from a third to 42\% of the party. They do not return to the pre-campaign values until 1959, showing that the effect of the campaign on party activity was sustained until long after the epidemic phase was over  \cite{Jeffreys:History}.  The reproduction potential, $R_p$  is well above the extinction threshold, Fig. \ref{politicalfig1.fig}(d), and above the epidemic threshold in 1947. The latter threshold rises due to the shrinking susceptible pool, but by 1960 is back below $R_p$, indicating that another period of growth may have naturally occurred in the 1960s due to shifting recycling of party members who left. It is more likely that political events in the 1960s initiated the post 1961 growth that did occur. The interpretation  of periodicity in membership numbers will be discussed in the conclusion. 

Not all the parameters could be estimated by data fitting, or from other known values. For example the parameter $g$, the fraction of recruits who become recruiting activists, was set at 0.5.  Alternative data fits were obtained for other values of $g=0.2$ and $g=1$. In the first case the number of activists in the party declined too quickly compared with the historical narratives or the time. In the later case the  activists rose  too high. Thus 0.5 is taken as a good compromise, with the model robust against small variations in this value. A similar argument can be applied to the initial number of activists $A_{T0}=I_0 + A_0$, and to time active $\tau_a$, neither of whose values are critical.

\subsection{UK Labour Party 1993--1997}\label{lab93.sec}
In 1992 the leadership of the Labour Party passed to John Smith who set about a recruitment campaign to rebuild the party in  a time of internal disillusionment, after a lengthy period out of electoral office \cite{Whiteley:High,Jeffreys:History}. Thus the start of the rise in party numbers is taken from 1993. The intention of Smith, and his successor from 1994 Tony Blair, was to increase membership in order to win the 1997 general election. This they succeeding in doing, with party membership rising to a peak in 1997, Table \ref{lab93.tab}.

As with Labour in the 1940s the initial population and open population are taken from the  electorate and voter base of the election prior to the simulation. The duration non-recruiting activist is set at 10 years for the same reasons as before. The fraction of activists is taken to be lower than the 1947 case, reflecting the poor state of engagement of the party membership in 1993 \cite{Whiteley:High,Jeffreys:History}.  The leaving rate is set at 15\% reflecting the greater nominal  party membership in the late 20th century, compared with the 1940s, and from the post 2000 decline. Membership of political parties has to be renewed annually, and thus people can leave by default unless they intentionally rejoin. The fraction recruited to the recruiting activists was taken to be lower than the 1947 case, $g=0.2$, as the party's intention was to increase its size to gain legitimacy for the 1997 election, rather than to build a permanent active membership \cite{Whiteley:High}.

\begin{figure}[!ht]
\begin{center}
\includegraphics[height=10cm]{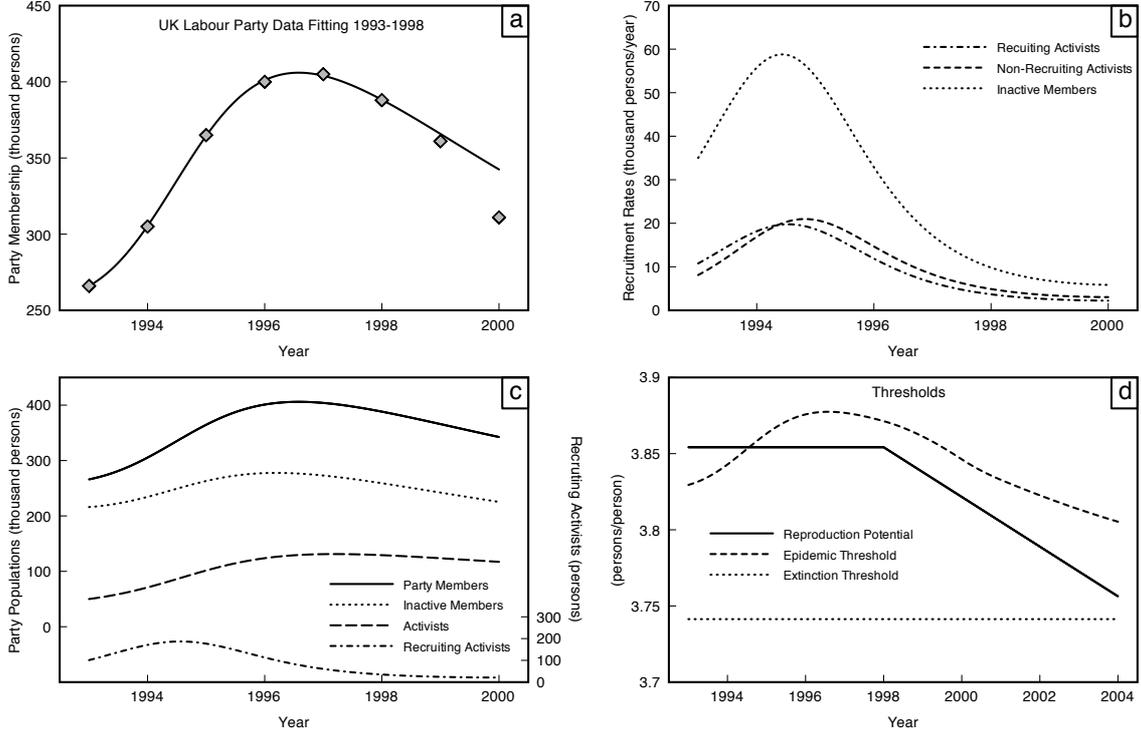}
\end{center}
\vspace{-18 pt}
\caption{Limited Activist Model applied to UK Labour Party. (a) Data fitting 1993--1998, showing deviation of model from post 1998 decline; (b) Recruitment rates to  party membership sub-populations $I$, $A$, $M$; (c) Party membership $P$ and sub-populations $I$, $I+A$, $M$; (d) Thresholds  $R_{\mathrm{epi}}$, $R_{\mathrm{ext}}$, compared with reproduction potential $R_p$, assuming falling $R_p$, and rising leaving rate $\alpha$, to account for post 1998 drop.}
\label{politicalfig2.fig}
\end{figure}

Data were fitted to the Limited Activist model from 1993--1998 (Table \ref{lab93.tab}), the latest year a good fit can be obtained, although the 1999 data point is also reasonable (Table \ref{lab3.tab}, Fig. \ref{politicalfig2.fig}(a)). The inactive members have the largest recruitment, in the early part of the campaign, with a more even balance later, Fig. \ref{politicalfig2.fig}(b). The recruiting activists peak before the end of 1994, not even rising to double their initial number, Fig. \ref{politicalfig2.fig}(c), yet this is sufficient to drive the large increase in the party and the total activists. The latter peak in 1997, the year of Labour's general election victory, Fig. \ref{politicalfig2.fig}(c), is consistent with  the known enthusiasm of the party at that time \cite{Whiteley:High}. The peak value of around a third of the party membership is  typical of party activism for a healthy party \cite{Janda:Political,Norris:Phoenix}.

\begin{table}[!ht] \footnotesize
\begin{center}
\begin{tabular} {lll}
  \hline
Estimated  Parameters & Value & Source \\
  \hline
Total  population $N$ & 43.25 mill  & UK electorate 1992 election \cite{uk:political}  \\
Open population $O=1-H$  &  11.56 mill & Labour voters 1992 election \cite{uk:political}\\
Initial party membership $P_0$ & 266,000& Party data \cite{keen2014membership} \\
Initial activists $ A_{T0}=I_0+A_0$ &50,000 & Estimate lower  than one third of party  active\\
&&    due to prior internal party conflict. \cite{Janda:Political,Norris:Phoenix}  \\
Duration non recruiting activist $\tau_{a}$ &10 years & Estimate similar to Labour 1940s\\
Leaving rate $\alpha$&0.15& Estimate from post 1997 decline\cite{keen2014membership}  \\
&& and presence of nominal members\cite{Whiteley:High}. \\
Fraction infected recruited $g$ & 0.2& Low ratio of active to inactive recruits \\ 
Recruitment potential $C_{p}$&19.2712 & \emph{Optimised}\\
Duration recruiting activist $\tau_{i}$& 0.00945 & \emph{Optimised}  \\
Initial recruiting Activists $ I_0$ &  101  & \emph{Optimised} \\
  \hline
\end{tabular}
\end{center}
\vspace{-10pt}
\caption{Parameter Values for Limited Activist Model of UK Labour Party Membership 1993--1998}
\label{lab3.tab}
\end{table}

The  epidemic threshold starts below the reproduction potential, initiating the rapid growth in the party, with growth ceasing as the threshold reaches its peak, Fig. \ref{politicalfig2.fig}(d). For fixed values of the recruitment potential $C_p$ and leaving rate $\alpha$ it is impossible to mimic the decline after 1998, the post election phase. Indeed had those values remained the party would have started growing again, in the early 2000s. However if the leaving rate is allowed to rise from 1998, and the recruitment potential to fall, perhaps due to the increasing focus of the party on government rather than retention and activism, then the data can be replicated to 2004. The reproduction  potential falls, staying under the epidemic threshold and approaching the extinction threshold, perhaps typical of a party giving little attention to its membership, Fig. \ref{politicalfig2.fig}(d).

\subsection{Scottish Nationalist Party (SNP) 2003--2013}

In 2003 the SNP failed to win power in Scotland for a second time in succession. In addition they lost members and chose to change leadership in order to rebuild the party. Alex Salmond became leader in 2004, and under his leadership the party membership increased by  70\% between 2003 and 2007, Table \ref{snp.tab}, achieving power as a minority government in 2007  \cite{uk:political,cairney2012scottish}.  With the need to gain legitimacy, and thus further members, for the 2011 Scottish parliament elections, the party saw another surge from 2010 to 2013, Table \ref{snp.tab}, winning the election outright. The data will be considered in these two natural phases, 2003-2010 and 2010-2013.

The initial population is taken from the typical Scottish electorate size 2003--2014, which remained reasonably stable throughout. Because of the increase in popularity of the SNP throughout the period, the open population is taken from an average of the SNP voters through all Scottish elections of the period. It is noted that fewer people vote SNP in UK general elections, which probably does not reflect the true sympathy for the party \cite{uk:political}. The duration non-recruiting activist  and  initial fraction activist are taken to be the same as the Labour party in the 1940s, as the SNP pursued a similarly positive recruitment drive \cite{snp:double}.   Likewise the fraction  recruited to recruiting activists, and the leaving rate, are taken to be the same as 1940s Labour. 

\begin{figure}[!ht]
\begin{center}
\includegraphics[height=10cm]{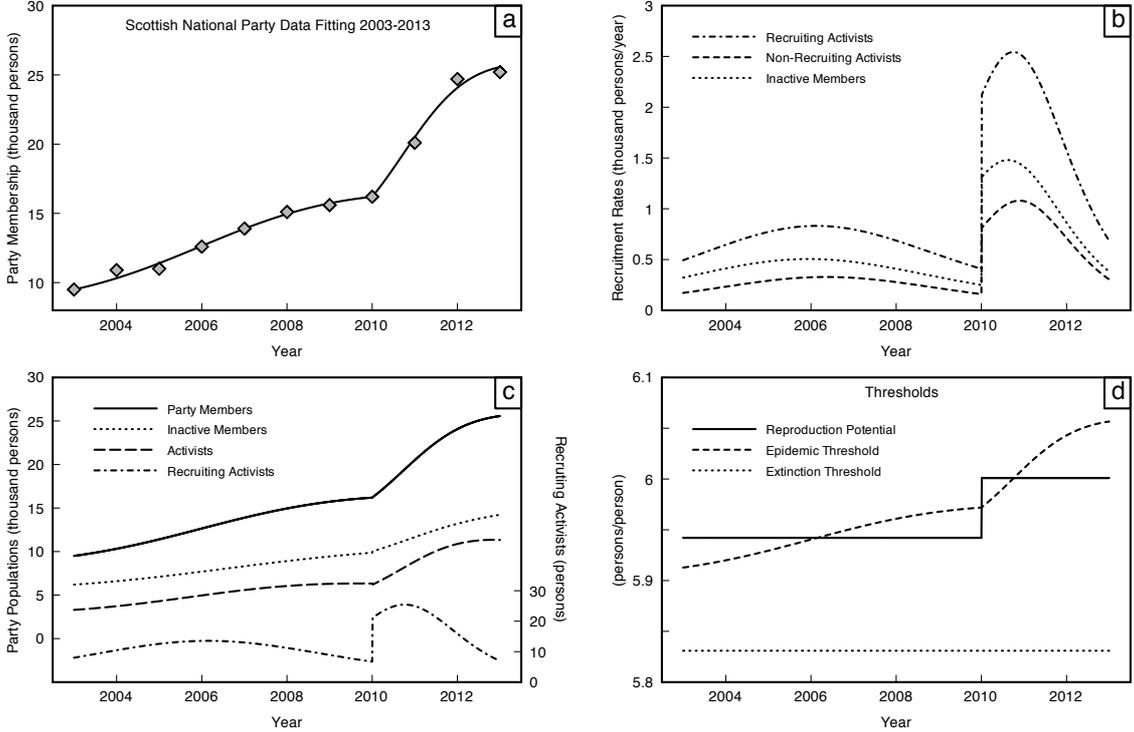}
\end{center}
\vspace{-18 pt}
\caption{Limited Activist Model applied to SNP. (a) Data fitting in two phases, 2003--2010 and 2010--2013; (b) Recruitment rates to  party membership sub-populations $I$, $A$, $M$; (c) Party membership $P$ and sub-populations $I$, $I+A$, $M$; (d) Thresholds  $R_{\mathrm{epi}}$, $R_{\mathrm{ext}}$, compared with reproduction potential $R_p$.}
\label{politicalfig3.fig}
\end{figure}

The Limited Activist model was fitted to data from 2003--2010 (the first phase) and then separately for 2010--2013 (the second phase), Table \ref{snp.tab}. The party numbers and number of activists were matched at 2010, Fig. \ref{politicalfig3.fig}(a--b), but a change in the initial recruiting activists was needed for the second phase, Table \ref{SNP1.tab}.  The recruiting activists have their first peak just before the Scottish election of 2007 and the party growth had almost finished by 2010, Fig. \ref{politicalfig3.fig}(b).

\begin{table}[!ht] \footnotesize
\begin{center}
\begin{tabular} {lll}
  \hline
Estimated  Parameters & Value & Source \\
  \hline
Total  population $N$ & 4 mill  & Scotland electorate  typical \\ && value 2003-2013 \cite{scot:elect}  \\
Open population $O=1-H$  &  686,000 & Average SNP voters 2003--2011\\&& Scottish elections \cite{uk:political}\\
Initial party membership $P_0$ & 9,500& Party data \cite{keen2014membership} \\
Initial activists $ A_{T0}=I_0+A_0$ &33,000 & Roughly a third of party active \cite{Janda:Political,Norris:Phoenix}  \\
Duration non recruiting activist $\tau_{a}$ &10 years & Estimate similar to Labour 1940s. \\
Leaving rate $\alpha$&0.05& Estimate similar to Labour 1940s.  \\
Fraction infected recruited $g$ & 0.5& Estimate similar to Labour 1940s.  \\ 
Recruitment potential $C_{p}$&(i) 11.8842, (ii) 12.002 & \emph{Optimised.}\\
Duration recruiting activist $\tau_{i}$& (i) 0.0163, (ii)  0.01 & \emph{Optimised.}  \\
Initial recruiting activists $ I_0$ &  (i) 8, (ii) 21  & \emph{Optimised.} \\
  \hline
\end{tabular}
\end{center}
\vspace{-10pt}
\caption{Parameter Values for Limited Activist Model of SNP Membership (i) 2003--2010, (ii) 2010--2013.}
\label{SNP1.tab}
\end{table}

In the second phase the party growth of 50\% in 3 years is explained by  a sudden jump in  the number, and effectiveness, of recruiting activists, Fig. \ref{politicalfig3.fig}(b,d), timed for the 2011 Scottish parliamentary elections.  The epidemic period ends early in the party growth, Fig. \ref{politicalfig3.fig}(d), with most of the party growth after this period. 

The model predicts that growth had finished by the time of the 2014 Scottish Independence Referendum campaign, Fig. \ref{politicalfig3.fig}(a), a long way short of the SNP's target of doubling their party membership before that date \cite{snp:double}. Subsequently, however, the party membership doubled within days of the referendum of 18/09/2014 which delivered a ``No'' result \cite{snp:monday}. The latest figure available,  92,200  \cite{snp:ninety}, shows an increase of five times the 2010 value which the model does not predict. This indicates a limitation in the applicability of the model, which will be discussed in the conclusion.

\subsection{UK Conservative Party 2005--2012}
The final application of the limited activism model is  the period of sustained decline in the UK Conservative Party that followed their brief rise from 2005 to 2006, Table \ref{con.tab}.  The hypothesis is explored that the decline is due to a large leaving rate with the recruitment potential remaining the same throughout the growing and declining period.

Due to the weakness of the party following a period in which they had lost two successive general elections, the duration non-recruiting activist is assumed to be shorter than the previous applications, $\tau_a=5$ years, and the recruitment to recruiting activists as being low, $g=0.2$. Likewise the initial number of activists is taken as significantly less than the one third used in the growing party scenarios of the SNP and Labour 1940s. The rapid decline from 2006 indicates a high leaving rate of about $\alpha = 20\%$.  Other population  parameters are set using the electorate and voter base, as before, Table \ref{con1.tab}.

\begin{figure}[!ht]
\begin{center}
\includegraphics[height=10cm]{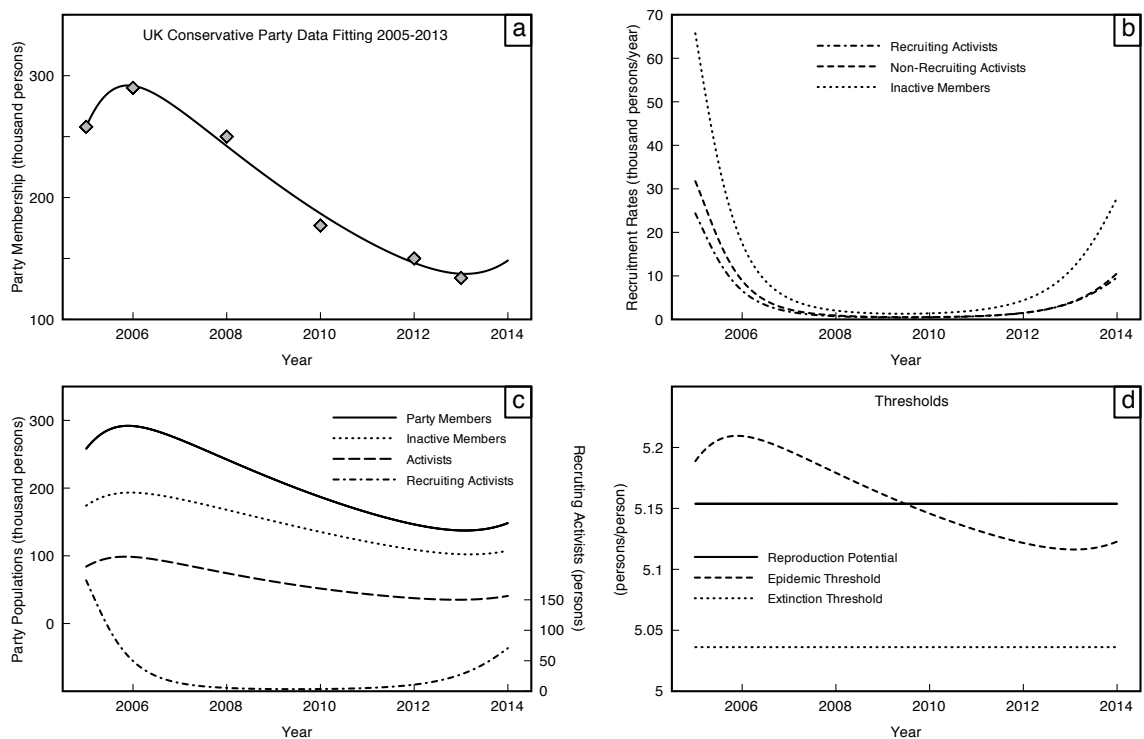}
\end{center}
\vspace{-18 pt}
\caption{Limited Activist Model applied to UK Conservative Party. (a) Data fitting 2005--2013, extrapolated to 2014; (b) Recruitment rates to  party membership sub-populations $I$, $A$, $M$; (c) Party membership $P$ and sub-populations $I$, $I+A$, $M$; (d) Thresholds $R_{\mathrm{epi}}$, $R_{\mathrm{ext}}$, compared with reproduction potential $R_p$.}
\label{politicalfig4.fig}
\end{figure}

The data fit is less precise than the other applications, Fig. \ref{politicalfig4.fig}(a). The reproduction potential is  lower than the epidemic threshold initially,  Fig. \ref{politicalfig4.fig}(d), indicating that the growth to 2006 was due to a large number of activists, rather than to their effectiveness in recruitment.  The subsequent decline is not only due to a high leaving rate but also due to virtually no recruitment 2007--2011, Fig. \ref{politicalfig4.fig}(b), which leads to a party that is proportionally more active due to the loss of inactive members, Fig. \ref{politicalfig4.fig}(c).

\begin{table}[!ht] \footnotesize
\begin{center}
\begin{tabular} {lll}
  \hline
Estimated  Parameters & Value & Source \\
  \hline
Total  population $N$ & 44.1802 mill  & UK electorate 2005 election \cite{scot:elect}  \\
Open population $O=1-H$  &  8.7726 mill & Conservative  voters 2005 UK general election \cite{uk:political}\\
Initial party membership $P_0$ &  258,000& Party data \cite{keen2014membership} \\
Initial activists $ A_{T0}=I_0+A_0$ &84,000 & Less than third of party active \cite{Janda:Political,Norris:Phoenix}  \\
&& a low value due to weakness of party.\\
Duration non recruiting activist $\tau_{a}$ &5 years & Estimate lower than previous data fits\\
&& due to party unpopularity. \\
Leaving rate $\alpha$&0.2& Estimate higher than previous data fits  \\
&& to account for rapid decline during period. \\
Fraction infected recruited $g$ & 0.2& Low ratio of active to inactive recruits. \\ 
Recruitment potential $C_{p}$&25.769 & \emph{Optimised.}\\
Duration recruiting activist $\tau_{i}$& 0.0074 & \emph{Optimised.}  \\
Initial recruiting Activists $ I_0$ &  182  & \emph{Optimised.} \\
  \hline
\end{tabular}
\end{center}
\vspace{-10pt}
\caption{Parameter Values for Limited Activist Model of UK Conservative Party  Membership 2005--2013}
\label{con1.tab}
\end{table}

The reproduction potential is above the extinction threshold, Fig. \ref{politicalfig4.fig}(d). Thus the model predicts a return to growth for the Conservative party from 2014. Although the party does not release official membership figures, a  tentative figure for 2014 of  149,800 \cite{con:home} is in line with the model's prediction. While there is a long-term gradual decline in political party membership \cite{Seyd:British,Whiteley:Dynamics}, the short-term rapid decline of the sort experienced by the Conservative party may be due to the dynamical effects of high leaving rates and moderate reproduction of recruiting activists.

\section{Conclusion}
The Limited Activist model  was developed to investigate the hypothesis that political parties grow through the action of activists persuading non members to join through word of mouth. The basic SIR construct gave a poor understanding of activists as ``infected'' spreaders of party membership due to the very short duration as an activist. However the revised format, where activists were compartmentalised into recruiters and non-recruiters was able to give convincing data fits for four different sets of data, with realistic durations for activists. Together with compartmentalising the non-members into those open to persuasion and those closed, each application of the model was able to give a narrative consistent with historical evidence for the party in question. Thus the epidemic analogy as an explanation of political party growth is strongly supported.

In all four application areas the reproduction potentials of the political parties were above the extinction threshold, as might be expected of parties who periodically achieve power. Despite party decline, neither of the three parties studied showed any indication that decline would lead to extinction, the DFE. Instead any decline comes from high reproduction rates with high leaving rates and the periodicity this induces. Such periodicity around the EE would not be expected to be perfect due to high variation in leaving rates from exogenous political factors such as elections and periods of political power. The decline of the Labour Party membership through the early 2000s could only be explained by such an increase in leaving rate, Section \ref{lab93.sec}. For constant rates the model explains a return to growth after a period of decline in terms of the recycling of members.  Given that in two of the applications studied the regrowth led to a return to political power for the party, it is suggested that electoral success may be connected with the natural recycling of members and its effect on activism.

In order for the concept of activists to have meaning in the model when only a fraction of them were recruiting, and thus contributing to growth, the model allowed the fraction of all activists, $f(A_t)$,  to influence the quality of recruitment.  Thus the more active the party, the more the  recruits became active participants themselves, a form of positive feedback. Improving active participation in the party is a key role of activists, more common than the level of recruitment itself \cite{Weldon:polity,Norris:Phoenix, pedersen2004sleeping}. In each of the four applications $f(A_t)$ had a significant variation. For Labour in the 1940s the fraction rose from a third to over 40\% with the fraction of the party active increasing substantially to 1952, Fig. \ref{politicalfig1.fig}(c). The value of $f(A_t)$ returned to a third by 1955, with the party becoming increasingly inactive. In the case of the SNP an increasing value of this fraction led to almost half the party active by 2012, Fig. \ref{politicalfig3.fig}{c}. Thus this modelling construct is believed to have added to the explanatory power of the model.

It is noted that in all four application areas  the duration recruiting activist was short, sometimes only a matter of days, and that at any given time the number of recruiting activists was low, less than 0.1\% of the party. Such figures are typical of word-of-mouth models applied to organisations with rapid growth, for example churches \cite{Hayward:Church}, in contrast to more individualised processes \citep[e.g.][]{Burbeck:Riot,Manthey:drink,gonzalez2003too} where the duration influential is longer. Political parties, like churches, have influential recruiters who only engage in recruitment activities  for a small portion of a month. \citet{Seyd:British} state that most activists devote no more than 10 hours a month  to party activities, which would mainly be organised regular party meetings. It would be expected that recruitment activities would likewise be conducted in small but repeated blocks, rather than one continuous period. Thus the short timescales of recruitment activity may  reflect the actual time spent recruiting with periods of inactivity removed as  aggregate models smooth the behaviour of different individuals over time. 

Although the model worked well over periods where there was a sustained recruitment campaign, or lack of one, it was not able to handle a sudden change in membership due to external political events. The massive increase in SNP membership in the days following the Scottish independence referendum  could not be predicted by the model. It is likely that a significant realignment towards the SNP took place during the 18-month long referendum campaign, but did not manifest itself in party membership until the referendum result was  known. It is possible that the ``No'' vote mobilised supporters of independence to join the SNP to continue the pressure for more devolved power for Scotland. Such a process could be modelled as a word-of-mouth process, where the closed non-member population in the Limited Activist model was allowed to change through the influence of supporters for independence. Such an extended  model would be more complex than the Limited Activist one presented, a complexity unnecessary for  most situations of party growth and decline.

The model could also be extended by allowing the activists to have a role in reactivating the long-term inactive members, rather than just the new recruits. Unlike recruitment campaigns it is not clear what the arena for such influence would be. Former active members could be contacted through friendship networks, however inactive members who had never engaged with the party may not be so easily contactable. Such an extension would require the disaggregation of the inactive members, raising further issues with estimating the ratio of  the  different types of new subpopulations and the extra parameters involved. 

The limitation of a purely population model of growth should be noted. The model limits the growth hypothesis to word of mouth. It is known that activists  also have a role in building the party's political legitimacy which in turn enhances growth \cite{Whiteley:High}. Such legitimacy also influences the engagement of party members, which in the model was approximated by the fraction of the party active $f(A_t)$. It would be interesting to explicitly include such legitimacy in the model as a variable, but this would be a non-population variable which would require care in its construction and use. Such an idea is reserved for a future publication.

It is proposed that the Limited Activist model put forward in this paper has sufficient richness to relate results to historical events, yet is simple enough to avoid over parametrisation. The model is recommended as useful first attempt in modelling political party growth  that researchers could use to explore and test on other data.

\section*{References}

\bibliographystyle{unsrtnat}   
\bibliography{JeffsHaywardPoliticalPartyGrowth}

\begin{thebibliography}{58}
\providecommand{\natexlab}[1]{#1}
\providecommand{\url}[1]{\texttt{#1}}
\expandafter\ifx\csname urlstyle\endcsname\relax
  \providecommand{\doi}[1]{doi: #1}\else
  \providecommand{\doi}{doi: \begingroup \urlstyle{rm}\Url}\fi

\bibitem[Michels(1966)]{Michels:Political}
R.~Michels.
\newblock \emph{Political parties: {A} sociological study of the oligarchical
  tendencies of modern democracy}.
\newblock Free Press, New York, 1966.

\bibitem[Tan(1998)]{Tan:size}
A.C. Tan.
\newblock The impacts of party membership size: A cross-national analysis.
\newblock \emph{Journal of Politics}, 60:\penalty0 188--198, 1998.

\bibitem[Olson(1971)]{Olson:Logic}
M.~Olson.
\newblock \emph{The logic of collective action: {P}ublic goods and the theory
  of groups}, volume 124.
\newblock Harvard University Press, Massachusetts, 1971.

\bibitem[Janda(1980)]{Janda:Political}
K.~Janda.
\newblock \emph{Political parties: {A} cross-national survey}.
\newblock Free Press, New York, 1980.

\bibitem[Seyd and Whiteley(2004)]{Seyd:British}
P.~Seyd and P.~Whiteley.
\newblock British party members: {A}n overview.
\newblock \emph{Party Politics}, 10\penalty0 (4):\penalty0 355--366, 2004.

\bibitem[Weldon(2006)]{Weldon:polity}
S.~Weldon.
\newblock Downsize my polity? {T}he impact of size on party membership and
  member activism.
\newblock \emph{Party Politics}, 12\penalty0 (4):\penalty0 467--481, 2006.

\bibitem[Whiteley(2009)]{Whiteley:Dynamics}
P.~Whiteley.
\newblock Where have all the members gone? {T}he dynamics of party membership
  in {B}ritain.
\newblock \emph{Parliamentary Affairs}, 62\penalty0 (2):\penalty0 242--257,
  2009.

\bibitem[Norris(2002)]{Norris:Phoenix}
P.~Norris.
\newblock \emph{Democratic phoenix: {R}einventing political activism}.
\newblock Cambridge University Press, Cambridge, 2002.

\bibitem[Tan(1997)]{Tan:decline}
A.C. Tan.
\newblock Party change and party membership decline: {A}n exploratory analysis.
\newblock \emph{Party Politics}, 3\penalty0 (3):\penalty0 363--377, 1997.

\bibitem[Dalton and Wattenberg(2000)]{Dalton:Partisans}
R.J. Dalton and M.P. Wattenberg.
\newblock \emph{Parties without partisans: political change in advanced
  industrial democracies}.
\newblock Oxford University Press, Oxford, 2000.

\bibitem[Whiteley and Seyd(2002)]{Whiteley:High}
P.~Whiteley and P.~Seyd.
\newblock \emph{High-intensity participation: {T}he dynamics of party activism
  in Britain}.
\newblock University of Michigan Press, Michigan, 2002.

\bibitem[Jefferys(2007)]{Jeffreys:History}
K.~Jefferys.
\newblock \emph{Politics and the people: {A} history of {B}ritish democracy
  since 1918}.
\newblock Atlantic Books, 2007.

\bibitem[Burbeck et~al.(1978)Burbeck, Raine, and Stark]{Burbeck:Riot}
S.L. Burbeck, W.J. Raine, and M.J.A. Stark.
\newblock The dynamics of riot growth: An epidemiological approach.
\newblock \emph{Journal of Mathematical Sociology}, 6\penalty0 (1):\penalty0
  1--22, 1978.

\bibitem[Hayward et~al.(2014)Hayward, Jeffs, Howells, and
  Evans]{hayward2014model}
J.~Hayward, R.A. Jeffs, L.~Howells, and K.S. Evans.
\newblock Model building with soft variables: A case study on riots.
\newblock In \emph{Proceedings of the 32nd International Conference of the
  System Dynamics Society}. System Dynamics Society Albany, NY, 2014.

\bibitem[Nizamani et~al.(2014)Nizamani, Memon, and Galam]{nizamani2013public}
S.~Nizamani, N.~Memon, and S.~Galam.
\newblock From public outrage to the burst of public violence: {A}n
  epidemic-like model.
\newblock \emph{Physica A: Statistical Mechanics and its Applications},
  416:\penalty0 620--630, 2014.

\bibitem[Camacho(2013)]{camacho2013development}
E.T. Camacho.
\newblock The development and interaction of terrorist and fanatic groups.
\newblock \emph{Communications in Nonlinear Science and Numerical Simulation},
  18\penalty0 (11):\penalty0 3086--3097, 2013.

\bibitem[Abrams and Strogatz(2003)]{Abrams:language}
D.M. Abrams and S.H. Strogatz.
\newblock Modelling the dynamics of language death.
\newblock \emph{Nature}, 424\penalty0 (6951):\penalty0 900, 2003.

\bibitem[Wyburn and Hayward(2008)]{wyburn2008future}
J.~Wyburn and J.~Hayward.
\newblock The future of bilingualism: An application of the {B}aggs and
  {F}reedman model.
\newblock \emph{Journal of Mathematical Sociology}, 32\penalty0 (4):\penalty0
  267--284, 2008.

\bibitem[Manthey et~al.(2008)Manthey, Aidoo, and Ward]{Manthey:drink}
J.L. Manthey, A.Y. Aidoo, and K.Y. Ward.
\newblock Campus drinking: an epidemiological model.
\newblock \emph{Journal of Biological Dynamics}, 2\penalty0 (3):\penalty0
  346--356, 2008.

\bibitem[S{\'a}nchez et~al.(2007)S{\'a}nchez, Wang, Castillo-Ch{\'a}vez,
  Gorman, and Gruenewald]{Sanchez:Drink}
F.~S{\'a}nchez, X.~Wang, C.~Castillo-Ch{\'a}vez, D.M. Gorman, and P.J.
  Gruenewald.
\newblock Drinking as an epidemic: A simple mathematical model with recovery
  and relapse.
\newblock In Witkiewitz K. and Marlatt A, editors, \emph{Therapist's Guide to
  Evidence based Relapse Prevention}, pages 353--368. Academic Press, New York,
  2007.

\bibitem[Rowe et~al.(1992)Rowe, Chassin, Presson, Edwards, and
  Sherman]{Rowe:Smoke}
D.~Rowe, L.~Chassin, C.~Presson, D.~Edwards, and S.~Sherman.
\newblock An ``epidemic'' model of adolescent cigarette smoking.
\newblock \emph{Journal of Applied Social Psychology}, 22:\penalty0 261--285,
  1992.

\bibitem[Gonzalez et~al.(2003)Gonzalez, Huerta-Sanchez, Ortiz-Nieves,
  Vazquez-Alvarez, and Kribs-Zaleta]{gonzalez2003too}
B.~Gonzalez, E.~Huerta-Sanchez, A.~Ortiz-Nieves, T.~Vazquez-Alvarez, and
  C.~Kribs-Zaleta.
\newblock Am {I} too fat? bulimia as an epidemic.
\newblock \emph{Journal of Mathematical Psychology}, 47\penalty0
  (5--6):\penalty0 515--526, 2003.

\bibitem[Santonja et~al.(2010)Santonja, Villanueva, J\'{o}dar, and
  Gonzalez-Parra]{santonja2010mathematical}
F.J. Santonja, R.J. Villanueva, L.~J\'{o}dar, and G.~Gonzalez-Parra.
\newblock Mathematical modelling of social obesity epidemic in the region of
  {V}alencia, {S}pain.
\newblock \emph{Mathematical and Computer Modelling of Dynamical Systems},
  16\penalty0 (1):\penalty0 23--24, 2010.

\bibitem[Kawachi(2008)]{Kawachi:Rumour}
K.~Kawachi.
\newblock Deterministic models for rumor transmission.
\newblock \emph{Nonlinear Analysis: Real World Applications}, 9:\penalty0
  1989--2028, 2008.

\bibitem[Zhao et~al.(2011)Zhao, Wang, Chen, Wang, Cheng, and Cui]{zhao2011sihr}
L.~Zhao, J.~Wang, Y.~Chen, Q.~Wang, J.~Cheng, and H.~Cui.
\newblock {SIHR} rumor spreading model in social networks.
\newblock \emph{Physica A: Statistical Mechanics and its Applications},
  391\penalty0 (7):\penalty0 2444--2453, 2011.

\bibitem[Santonja et~al.(2008)Santonja, Tarazona, and
  Villanueva]{santonja2008mathematical}
F.J. Santonja, A.C. Tarazona, and R.J. Villanueva.
\newblock A mathematical model of the pressure of an extreme ideology on a
  society.
\newblock \emph{Computers \& Mathematics with Applications}, 56\penalty0
  (3):\penalty0 836--846, 2008.

\bibitem[Vitanov et~al.(2010)Vitanov, Dimitrova, and
  Ausloos]{vitanov2010verhulst}
N.K. Vitanov, Z.I. Dimitrova, and M.~Ausloos.
\newblock Verhulst-lotka-volterra ({VLV}) model of ideological struggle.
\newblock \emph{Physica A: Statistical Mechanics and its Applications},
  389\penalty0 (21):\penalty0 4970--4980, 2010.

\bibitem[Cannarella and Spechler(2014)]{cannarella2014epidemiological}
J.~Cannarella and J.A. Spechler.
\newblock Epidemiological modeling of online social network dynamics.
\newblock \emph{arXiv preprint arXiv:1401.4208}, 2014.

\bibitem[Woo et~al.(2013)Woo, Lee, Ku, and Chen]{woo2013modeling}
J.~Woo, M.J. Lee, Y.~Ku, and H.~Chen.
\newblock Modeling the dynamics of medical information through web forums in
  medical industry.
\newblock \emph{Technological Forecasting and Social Change}, 2013.

\bibitem[Chhabra et~al.(2014)Chhabra, Brundavanam, and
  Shannigrahi]{chhabra2014alternative}
S.S. Chhabra, A.~Brundavanam, and S.~Shannigrahi.
\newblock An alternative explanation for the rise and fall of {M}y{S}pace.
\newblock \emph{arXiv preprint arXiv:1403.5617}, 2014.

\bibitem[Bettencourt et~al.(2006)Bettencourt, Cintr{\'o}n-Arias, Kaiser, and
  Castillo-Ch{\'a}vez]{bettencourt2006power}
L.~Bettencourt, A.~Cintr{\'o}n-Arias, D.I. Kaiser, and C.~Castillo-Ch{\'a}vez.
\newblock The power of a good idea: Quantitative modeling of the spread of
  ideas from epidemiological models.
\newblock \emph{Physica A: Statistical Mechanics and its Applications},
  364:\penalty0 513--536, 2006.

\bibitem[Hayward(1999)]{Hayward:Church}
J.~Hayward.
\newblock Mathematical modeling of church growth.
\newblock \emph{Journal of Mathematical Sociology}, 23\penalty0 (4):\penalty0
  255--292, 1999.

\bibitem[Hayward(2005)]{Hayward:General}
J.~Hayward.
\newblock A general model of church growth and decline.
\newblock \emph{Journal of Mathematical Sociology}, 29\penalty0 (3):\penalty0
  177--207, 2005.

\bibitem[Ochoche and Gweryina(2013)]{ochoche2013evaluating}
J.M. Ochoche and R.I. Gweryina.
\newblock Evaluating the impact of active members on church growth.
\newblock \emph{International Journal of Science and Technology}, 2\penalty0
  (11), 2013.

\bibitem[Madubueze and Nwaokolo(2014)]{madubueze2014mathematical}
C.E. Madubueze and M.A. Nwaokolo.
\newblock A mathematical model to study the effect of renewal and reversion of
  inactive {C}hristians on church growth.
\newblock \emph{International Journal of Science and Technology}, 14\penalty0
  (2), 2014.

\bibitem[McCartney and Glass(2015)]{mccartney2015three}
M.~McCartney and D.H. Glass.
\newblock A three-state dynamical model for religious affiliation.
\newblock \emph{Physica A: Statistical Mechanics and its Applications},
  419:\penalty0 145--152, 2015.

\bibitem[Ausloos and Petroni(2007)]{ausloos2007statistical}
M.~Ausloos and F.~Petroni.
\newblock Statistical dynamics of religions and adherents.
\newblock \emph{EPL (Europhysics Letters)}, 77\penalty0 (3):\penalty0 38002,
  2007.

\bibitem[Ausloos and Petroni(2009)]{ausloos2009statistical}
M.~Ausloos and F.~Petroni.
\newblock Statistical dynamics of religion evolutions.
\newblock \emph{Physica A: Statistical Mechanics and its Applications},
  388\penalty0 (20):\penalty0 4438--4444, 2009.

\bibitem[Ausloos(2012)]{ausloos2012econophysics}
M.~Ausloos.
\newblock Econophysics of a religious cult: the {A}ntoinists in {B}elgium
  [1920--2000].
\newblock \emph{Physica A: Statistical Mechanics and its Applications},
  391\penalty0 (11):\penalty0 3190--3197, 2012.

\bibitem[Ausloos and Petroni(2010)]{Ausloos:econ}
M.~Ausloos and F.~Petroni.
\newblock On world religion adherence distribution evolution.
\newblock In M.~Takayasu, T.~Watanabe, and H.~Takayasu, editors,
  \emph{Econophysics Approaches to Large-Scale Business Data and Financial
  Crisis}, pages 289--312. Springer, 2010.

\bibitem[Abrams et~al.(2011)Abrams, Yaple, and Wiener]{abrams2011dynamics}
D.M. Abrams, H.A. Yaple, and R.J. Wiener.
\newblock Dynamics of social group competition: modeling the decline of
  religious affiliation.
\newblock \emph{Physical {R}eview {L}etters}, 107\penalty0 (8):\penalty0
  088701, 2011.

\bibitem[Romero et~al.(2011)Romero, Kribs-Zaleta, Mubayi, and
  Orbe]{romero2009epidemiological}
D.M. Romero, C.M. Kribs-Zaleta, A.~Mubayi, and C.~Orbe.
\newblock An epidemiological approach to the spread of political third parties.
\newblock \emph{Discrete and Continuous Dynamical Systems-Series B (DCDS-B)},
  15\penalty0 (3):\penalty0 707--738, 2011.

\bibitem[Galam and Wonczak(2000)]{galam2000dictatorship}
S.~Galam and S.~Wonczak.
\newblock Dictatorship from majority rule voting.
\newblock \emph{The European Physical Journal B-Condensed Matter and Complex
  Systems}, 18\penalty0 (1):\penalty0 183--186, 2000.

\bibitem[Qian et~al.(2015)Qian, Liu, and Galam]{qian2015activeness}
S.~Qian, Y.~Liu, and S.~Galam.
\newblock Activeness as a key to counter democratic balance.
\newblock \emph{Physica A: Statistical Mechanics and its Applications},
  432:\penalty0 187--196, 2015.

\bibitem[Hethcote et~al.(2002)Hethcote, Zhien, and Shengbing]{hethcote:effects}
H.~Hethcote, M.~Zhien, and L.~Shengbing.
\newblock Effects of quarantine in six endemic models for infectious diseases.
\newblock \emph{Mathematical biosciences}, 180\penalty0 (1):\penalty0 141--160,
  2002.

\bibitem[Anderson and May(1992)]{anderson1992infectious}
R.M. Anderson and R.M. May.
\newblock \emph{Infectious diseases of humans: dynamics and control},
  volume~26.
\newblock Oxford University Press, Oxford, 1992.

\bibitem[uk:(2012)]{uk:political}
{U}{K} {P}olitical {I}nfo, 2012.
\newblock URL \url{http://www.ukpolitical.info/}.
\newblock {L}ast accessed 1st March 2015.

\bibitem[Keen(2014)]{keen2014membership}
R.~Keen.
\newblock Membership of {UK} political parties.
\newblock Research briefing: standard notes SN/SG/5125, House of Commons,
  London, UK, 2014.

\bibitem[Pedersen et~al.(2004)Pedersen, Bille, Buch, Elklit, Hansen, and
  Nielsen]{pedersen2004sleeping}
K.~Pedersen, L.~Bille, R.~Buch, J.~Elklit, B.~Hansen, and H.~J. Nielsen.
\newblock Sleeping or active partners? {D}anish party members at the turn of
  the millennium.
\newblock \emph{Party Politics}, 10\penalty0 (4):\penalty0 367--383, 2004.

\bibitem[Granovetter(1978)]{granovetter1978threshold}
M.~Granovetter.
\newblock Threshold models of collective behavior.
\newblock \emph{American journal of sociology}, pages 1420--1443, 1978.

\bibitem[Ausloos(2015)]{ausloos2015slow}
M.~Ausloos.
\newblock Slow-down or speed-up of inter- and intra-cluster diffusion of
  controversial knowledge in stubborn communities based on a small world
  network.
\newblock \emph{Frontiers in Physics}, 3, 2015.
\newblock \doi{10.3389/fphy.2015.00043}.

\bibitem[Cairney(2012)]{cairney2012scottish}
P.~Cairney.
\newblock \emph{The Scottish Political System Since Devolution: From New
  Politics to the New Scottish Government}.
\newblock Andrews UK Limited, 2012.

\bibitem[snp(2011)]{snp:double}
{S}{N}{P} set membership challenge.
\newblock SNP, website, October 17, 2011. \newline
\newblock URL
  \url{http://www.snp.org/media-centre/news/2011/oct/snp-set-membership-challenge}. \newline
\newblock {L}ast accessed 1st March 2015.

\bibitem[sco(2012)]{scot:elect}
Electoral statistics - {S}cotland 1st {D}ecember 2011.
\newblock General Register Office of Scotland, website, February 22 2012.\newline
\newblock URL
  \url{http://www.gro-scotland.gov.uk/statistics-and-data/statistics/statistics-by-theme/} \newline
  \newblock
  \url{electoral-statistics/1st-december-2011}. \newline
\newblock {L}ast accessed 11th September 2015.

\bibitem[snp(2014{\natexlab{a}})]{snp:monday}
{S}{N}{P} becomes uk's third biggest party in wake of indyref defeat.
\newblock SNP, website, September 22, 2014{\natexlab{a}}.\newline
\newblock URL
  \url{http://www.heraldscotland.com/politics/referendum-news/}\newline
  \newblock 
 \url{   snp-on-course-to-be-uks-third-biggest-political-party.25402590}.\newline
\newblock {L}ast accessed 1st March 2015.

\bibitem[snp(2014{\natexlab{b}})]{snp:ninety}
{S}{N}{P} membership now exceeds extraordinary 90,000.
\newblock SNP, website, November 22, 2014{\natexlab{b}}.
\newblock URL
  \url{http://www.snp.org/media-centre/news/2014/nov/snp-membership-now-exceeds-extraordinary-90000}.
\newblock {L}ast accessed 1st March 2015.

\bibitem[con(2014)]{con:home}
{C}onservative {P}arty membership has risen to 149,800 - up 11.7 per cent.
\newblock Conservative Party, website, September 28, 2014.
\newblock URL
  \url{http://www.conservativehome.com/thetorydiary/2014/09/conference-survey-and-membership-figures.html}.
\newblock {L}ast accessed 1st March 2015.

\bibitem[new(2012)]{new:state}
How {T}ory membership has collapsed under {C}ameron.
\newblock New Statesman, website, July 30, 2012.
\newblock URL
  \url{http://www.newstatesman.com/blogs/politics/2012/07/how-tory-membership-has-collapsed-under-cameron}.
\newblock {L}ast accessed 1st March 2015.

\end{thebibliography}

\appendix
\section{Party Membership Data}

\begin{table}[!ht] \footnotesize
\begin{center}
\begin{tabular} {*{12}{c}}
  \hline
  Year & 1944 & 1945 & 1946 & 1947 & 1948  & 1949 & 1950 & 1951 & 1952 & 1953\\
  \hline
 Membership & 266 & 487 & 645 & 608 & 629 &  730 & 908 & 876 & 1,015 & 1,005 \\
  \hline
\end{tabular}
\end{center}
\vspace{-10pt}
\caption{Labour Party Membership 1944-1953 (000s) \cite{keen2014membership}}
\label{lab44.tab}
\end{table}

\begin{table}[!ht] \footnotesize
\begin{center}
\begin{tabular} {*{13}{c}}
  \hline
  Year & 1993 &1994 & 1995 & 1996 & 1997 & 1998 & 1999 & 2000 & 2001 & 2002 & 2003 & 2004 \\
  \hline
 Membership & 266 & 305 & 365 & 400 & 405 & 388 & 361 & 311 & 272 & 248 & 215 & 198 \\
  \hline
\end{tabular}
\end{center}
\vspace{-10pt}
\caption{Labour Party Membership 1993-2004 (000s) \cite{keen2014membership}}
\label{lab93.tab}
\end{table}

\begin{table}[!ht] \footnotesize
\begin{center}
\begin{tabular} {*{13}{c}}
  \hline
  Year & 2003 & 2004 & 2005 & 2006 & 2007  & 2008 & 2009 & 2010 & 2011 & 2012 & 2013 \\
  \hline
 Membership & 9.5 & 10.9 & 11.0 & 12.6 & 13.9 &  15.1 & 15.6 & 16.2 & 20.1 & 24.7  & 25.2  \\
  \hline
\end{tabular}
\end{center}
\vspace{-10pt}
\caption{Scottish Nationalist Party Membership 2003-2013  (000s)  \cite{keen2014membership}}
\label{snp.tab}
\end{table}

\begin{table}[!ht] \footnotesize
\begin{center}
\begin{tabular} {*{7}{c}}
  \hline
  Year & 2005 & 2006 &2008 & 2010 & 2012  & 2013  \\
  \hline
 Membership & 258 & 290 & 250 & 177 & 150  & 134 \\
  \hline
\end{tabular}
\end{center}
\vspace{-10pt}
\caption{Conservative Party Membership 2005--2013 (000s) \cite{keen2014membership,new:state}}
\label{con.tab}
\end{table}

\section{Proof $A_{+}<0$ given $\hat{p}>0$}\label{proof.appendix}
If $A_+ = \hat{p}-\hat{q}\bar{I_{+}}<0$ then using (\ref{model.a1}):
\begin{equation}
    2a_1-\hat{q}\left(b_1/\hat{p} + \sqrt{ (b_1/\hat{p})^2 - 4a_1c_1/\hat{p}^2} \right)< 0 \label{model.ee5}
  \end{equation}
  as $a_1>0$ and $\hat{p}>0$
  
  (\ref{model.ee5}) can be re-arranged as:
\begin{equation}
       2a_1-\hat{q}b_1/\hat{p} < \hat{q}\sqrt{ (b_1/\hat{p})^2 - 4a_1c_1/\hat{p}^2}\label{model.ee7}
\end{equation}

There are two options, either $2a_1 < \hat{q}b_1/\hat{p}$ in which case the left hand side of (\ref{model.ee7}) is negative and  the result  immediately follows, or $2a_1 > \hat{q}b_1/\hat{p}$. In the latter case (\ref{model.ee7}) can be squared as both sides are positive:  $ \left(2a_1 -\hat{q}b_1/\hat{p}\right)^2     <\hat{q}^2 \left( (b_1/\hat{p})^2 -  4a_1c_1/\hat{p}^2 \right)$. This simplifies to $0 <g\alpha\tau_i+g $, thus proved.
 
\end{document}